\documentclass[twocolumn, 9pt]{extarticle}
\usepackage{amsmath}
\usepackage{lineno}
\usepackage[table,x11names,dvipsnames,table]{xcolor}
\usepackage{authblk}
\usepackage{subcaption,booktabs}
\usepackage{graphicx}
\usepackage{multirow}
\usepackage[nolist,nohyperlinks]{acronym}
\usepackage{xurl}
\usepackage{hyperref}
\usepackage[authoryear,square,semicolon]{natbib}
\usepackage{tabularx}
\usepackage{float}
\usepackage[group-separator={,}]{siunitx}
\usepackage{geometry}
\usepackage{comment}

\newcommand\B{\rule[-2.4ex]{0pt}{0pt}} 

\makeatletter
\def\@maketitle{
\raggedright
\newpage
  \noindent
  \vspace{0cm}
  \let \footnote \thanks
    {\hskip -0.4em \huge \textbf{{\@title}} \par}
    \vskip 1.5em
    {\large
      \lineskip .5em
      \begin{tabular}[t]{l}
      \raggedright
        \@author
      \end{tabular}\par}
    \vskip 1em
  \par
  \vskip 1.5em
  }
\makeatother

\begin{document}

\title{Communicating the gravitational-wave discoveries of the \\LIGO–Virgo–KAGRA Collaboration}

\author{Hannah Middleton\thanks{hannah.middleton@ligo.org}}
\author{Christopher~P~L~Berry\thanks{christopher.berry@ligo.org}} 
\author{Nicolas~Arnaud}
\author{David~Blair}
\author{Jacqueline~Bondell}
\author{Alice~Bonino}
\author{Nicolas~Bonne}
\author{Debarati~Chatterjee}
\author{Sylvain~Chaty}
\author{Storm~Colloms}
\author{Lynn~Cominsky} 
\author{Livia~Conti}
\author{Isabel~Cordero-Carrión}
\author{Robert~Coyne}
\author{Zoheyr~Doctor}
\author{Andreas~Freise}
\author{Aaron~Geller}
\author{Anna~C~Green}
\author{Jen~Gupta}
\author{Daniel~Holz}
\author{William~Katzman}
\author{Jyoti~Kaur}
\author{David~Keitel}
\author{Joey~Shapiro~Key}
\author{Nutsinee~Kijbunchoo}
\author{Carl~Knox}
\author{Coleman~Krawczyk}
\author{Ryan~N~Lang} 
\author{Shane~L~Larson}
\author{Susanne~Milde}
\author{Vincenzo~Napolano}
\author{Chris~North}
\author{Sascha~Rieger}
\author{Giada~Rossi}
\author{Hisaaki~Shinkai}
\author{Aurore~Simonnet}
\author{Andrew~Spencer}

\affil[1]{Author biographies and affiliations are included at the end of the article.}

\setcounter{Maxaffil}{0}
\renewcommand\Affilfont{\itshape\small}

\date{}  
\maketitle


\begin{abstract}
The LIGO–Virgo–KAGRA (LVK) Collaboration has made breakthrough discoveries in gravitational-wave astronomy, a new field that provides a different means of observing our Universe. Gravitational-wave discoveries are possible thanks to the work of thousands of people from across the globe working together. In this article, we discuss the range of engagement activities used to communicate LVK gravitational-wave discoveries and the stories of the people behind the science, using the activities surrounding the release of the third Gravitational-Wave Transient Catalog as a case study.
\end{abstract}

\section{Introduction: Context and Objectives}

Gravitational waves (GWs) can be difficult to imagine. 
They are ripples in spacetime, created by the acceleration of massive objects, that propagate across the Universe at the speed of light.  
The strongest GW sources are massive and rapidly moving: the best sources for ground-based GW detectors are coalescing binaries of black holes or neutron stars. 
After travelling across the cosmos to Earth, GWs are almost imperceptibly small. 
The strongest signals correspond to a stretching and squeezing of space of $1$ part in $10^{21}$. 
Observing GWs has therefore been a great experimental challenge, requiring an international community to design and build the complex, highly sensitive instruments needed to measure the waves, and to create and test the algorithms needed to analyse the data.

In 2015, $100$ years after Einstein first calculated the properties of GWs in his theory of general relativity, the first direct observation of GWs was made~\citep{AbbottEtAl:2016}. 
This discovery was confirmation for Einstein's theory, an experimental triumph, and the beginning of a new era for astronomy. 
The signal GW150914 came from two black holes, each about $30$ times the mass of our Sun, coalescing more than a billion light-years away. 
This was the first time such a binary black hole system was found, the first time black holes of this size were discovered, and the first time two black holes were observed to inspiral and merge. 
The discovery was a major global news event~\citep{KeyHendryHolz:2016}, building on the public’s fascination with Einstein and black holes. 

Since 2015, the LIGO–Virgo–KAGRA (LVK) Collaboration, which operates the international network of ground-based laser-interferometric GW observatories (the two LIGO sites in the US~\citep{LIGOAasiEtAl:2015}, Virgo in Italy~\citep{VirgoAcerneseEtAl:2015}, KAGRA in Japan~\citep{KAGRAAkutsuEtAl:2021} and GEO\,600 in Germany~\citep{GEO600DooleyEtAl:2016} has made many further GW discoveries. 
The field has grown rapidly, with the third GW Transient Catalog (GWTC-3) increasing the number of probable detections to $90$~\citep{AbbottEtAlGWTC3:2023}. 
The latest observing run is ongoing. 
These observations have revealed a diverse range of black hole and neutron star binaries, and revolutionised our understanding of these sources’ astrophysics.

Surrounding each results release are numerous engagement and communication activities. 
These share the wonder of discovery, what can be learned from GWs, and the technological advances that enable these groundbreaking measurements. 
Matching the global composition of the LVK, selected resources have been translated into over $20$ languages.

Engagement, communication and education activities are carried out by a wide variety of individuals: from those who see themselves as scientists and engineers with a passion for outreach, to those who view themselves as outreach experts with a passion for science. 
These individuals make commitments as LVK members, and join together in collaboration-wide teams and outreach partnerships.

Discoveries also provide an opportunity to focus on the people behind the science. 
The LVK includes around $2000$ people from diverse backgrounds across the globe. 
Their activities range from designing instrumentation to operating the kilometre-scale laser interferometers, from coding analysis algorithms to calculating GW emission. 
Making connections with the people within the LVK reveals the human side of how science works, and provides possible role models for young people interested in science, technology, engineering and mathematics (STEM).

Communicating LVK discoveries faces several challenges. Some, such as explaining complicated ideas using non-technical language or engaging with hard-to-reach communities, are common to other science communications. 
However, LVK science also faces specific challenges including explaining the concept of GWs, which are alien to everyday experiences; constructing a narrative around the work of a large collaboration, where it is often not possible to identify distinct contributions of individuals, and maintaining interest as the field develops from an era of first detections to using large catalogs to make statistical statements. 
Each of these challenges also provides an opportunity, enabling us to increase the scientific literacy of those reached: explaining the results of LVK work, how they are produced, and how modern science makes discoveries through carefully combining many pieces of research.

In this article, we detail communication methods used by the LVK. 
We first review the range of engagement activities used to communicate LVK GW discoveries and the stories of the people behind the science, and then provide specific examples of activities used for GWTC-3. 
We conclude with reflections upon these activities and the challenges of communicating LVK discoveries.

\section{General communication methods}

\begin{table*}
\begin{center}
\caption{\label{tab:activities}
A selection of LVK communication activities and examples we discuss in this article.
}
\begin{tabular}{p{0.20\textwidth} p{0.22\textwidth} p{0.22\textwidth} p{0.22\textwidth}}
\toprule 
\textbf{Activity} & \textbf{Examples} & \textbf{Audiences} & \textbf{Media} \\
\midrule 
Sharing the people behind the science & 
LIGO Magazine; Humans of LIGO; Antimatter comics; LIGO-India blog & 
LVK members; other academics; undergraduate students; school students; general public \B & 
Print; online writing; graphics  \\

Engaging academia &
Journal articles; Open Data Workshops; data releases; webinars; direct interventions &
LVK members; other academics; undergraduate students; interested public &
Academic resources; live and recorded talks; online course materials; online documentation; online correspondence \B \\

Supporting formal classroom education & 
Educator’s guide; Einstein-First; Space Public Outreach Team \B & 
School educators; undergraduate students; school children & 
Online written resources; face-to-face communication \\

Writing reference texts & 
Science Summaries; press releases; news items & 
Undergraduate students; science journalists; interested public \B & 
Online writing; print\\

Interacting in person & 
Science festivals; museum exhibits; visitor centres & 
School children; general public & 
Face-to-face communication; talks; live demonstrations; interactive exhibits \B \\

Creating graphics & 
Infographics; simulation visualisations; artistic impressions; Masses in the Stellar Graveyard; LVK Orrery \B & 
Academics; undergraduate students; journalists; general public & 
Images; videos \\[0.7cm]

Employing multisensory resources \B & 
Sounds of Spacetime; Tactile Universe; Low Mass Beats & 
General public & 
Audio; 3D-printed materials \\

Combining art and music with science & 
GravitySynth; GWSciArt; Celebrating Einstein festival \B & 
General public & 
Visual arts; music \\

Communicating through social and non-traditional media \B & 
Podcasts; social-media posting; Reddit Ask-me-anything & 
Science journalists; interested public; general public & 
Online correspondence; online writing; audio; graphics; videos\\

Using interactive technologies & 
Laser Labs games and apps; Black Hole Hunter; Mission Gravity & 
School children; general public & 
Interactive software \\

\bottomrule
\end{tabular}
\end{center}
\end{table*}

Discovery announcements from the LVK normally have three components: paper, data release, and educational and engagement resources. 
Here we describe some general and long-term education and engagement activities of the LVK.

The LVK is a consortium of its three constituent collaborations, each with its own internal organisational structure. 
Communications efforts are primarily led by a division-level coordinator or chair, who oversees various working groups, committees, and individual efforts working towards shared objectives. 
Activities organised through these groups are typically undertaken by teams of individuals, who then report their progress through regular meetings, LVK-wide conferences, and annual reports. 
Some activities, like coordination of press releases, may be done by dedicated communications teams in each of the three collaborations, but most activities are done by volunteers from the LVK. 
Division leads are responsible for ensuring that these activities complement each other and align with parallel efforts across the observatory sites.

The LVK consists of many individuals with different backgrounds and job roles working in institutions around the world and funded by different sources. 
This results in a diverse set of public-engagement activities run by different teams. 
While these may share a broad goal of communicating GW science, they may differ in specific goals, e.g., encouraging schoolchildren to study STEM or informing science journalists. The different activities complement each other, and provide a breadth of resources and expertise for LVK members to draw upon. 
Below we group in terms of activities (Table~\ref{tab:activities}), reflecting how these are organised within the LVK by different teams. 

\subsection{Sharing the people behind the science}

Discovery science is not just about the results, it is also about the people behind them. 
Highlighting individuals within the LVK can provide STEM role models, challenge stereotypes, and help to foster a welcoming environment by promoting diversity and inclusion.

The LIGO Magazine (\href{https://www.ligo.org/magazine}{www.ligo.org/magazine}), published twice annually, is commissioned and edited by a team of volunteers from the LIGO Scientific Collaboration. 
It was conceived to build connections across the diverse GW community and inform the interested public. Since the first edition in 2012, the Magazine has featured 470 authors, representing $237$ workplaces in $23$ countries (as of Issue 24, March 2024). 
Articles include perspectives of working on GW science, personal stories, opinion pieces, science explainers, and advice columns. 
The Magazine provides opportunities for early-career scientists to discuss their behind-the-scenes experiences of working on big GW results, as well as an avenue for current and former LVK members to share their perspectives of working in academia and beyond. 
The LIGO Magazine has a professional, high-quality design. 
For each edition, $500$ hard copies are printed. 
Approximately half are sent to LVK Collaboration meetings for participants to take back to their home institutions; the rest are distributed between detector sites and other LVK institutions worldwide. 
Print copies are also used within LVK as handouts for government representatives, students, and to visitors at detector sites and institutions. 
The full archive is also online as PDF downloads.

\begin{figure}
\centering
\includegraphics[width=.5\textwidth]{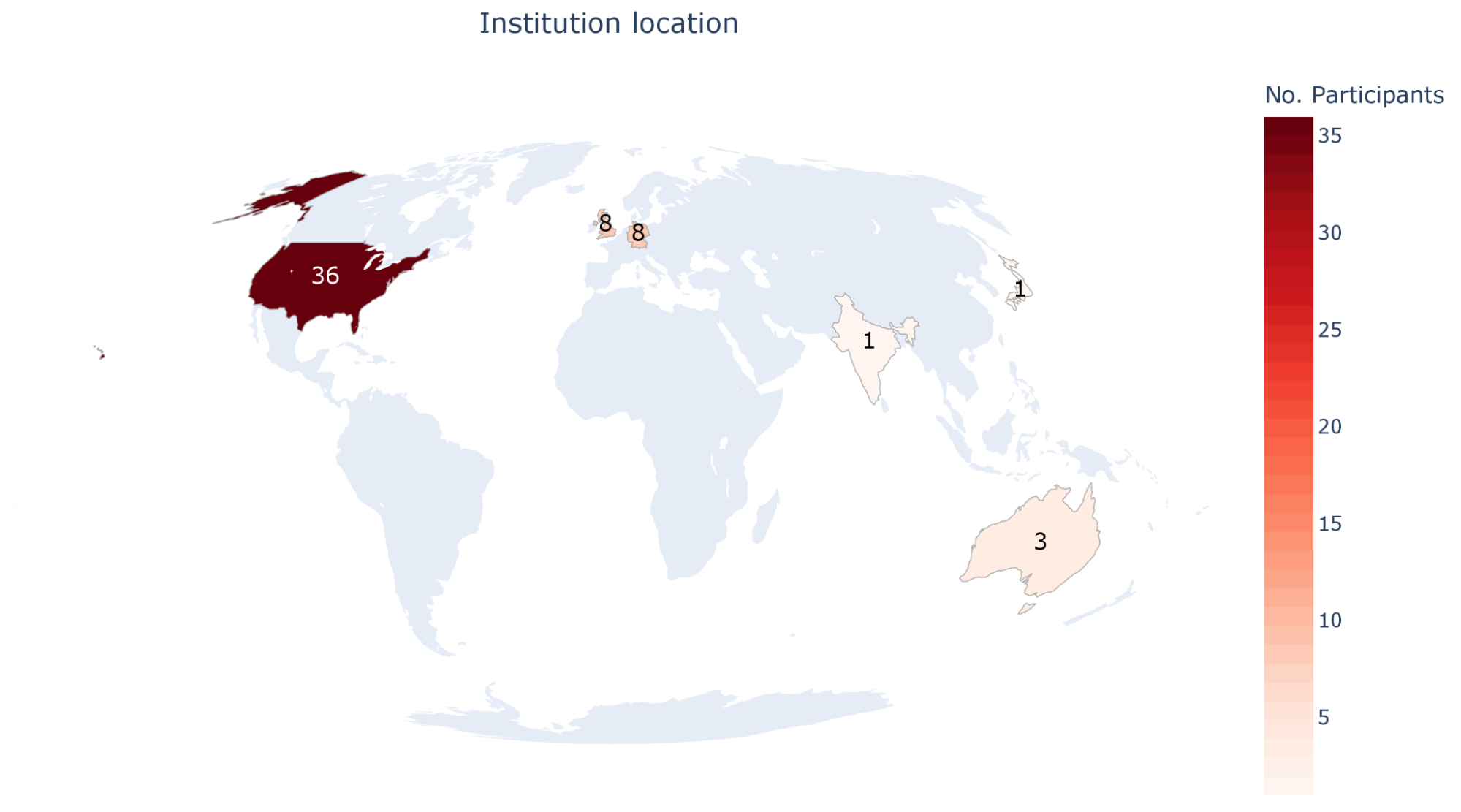}
\includegraphics[width=.5\textwidth]{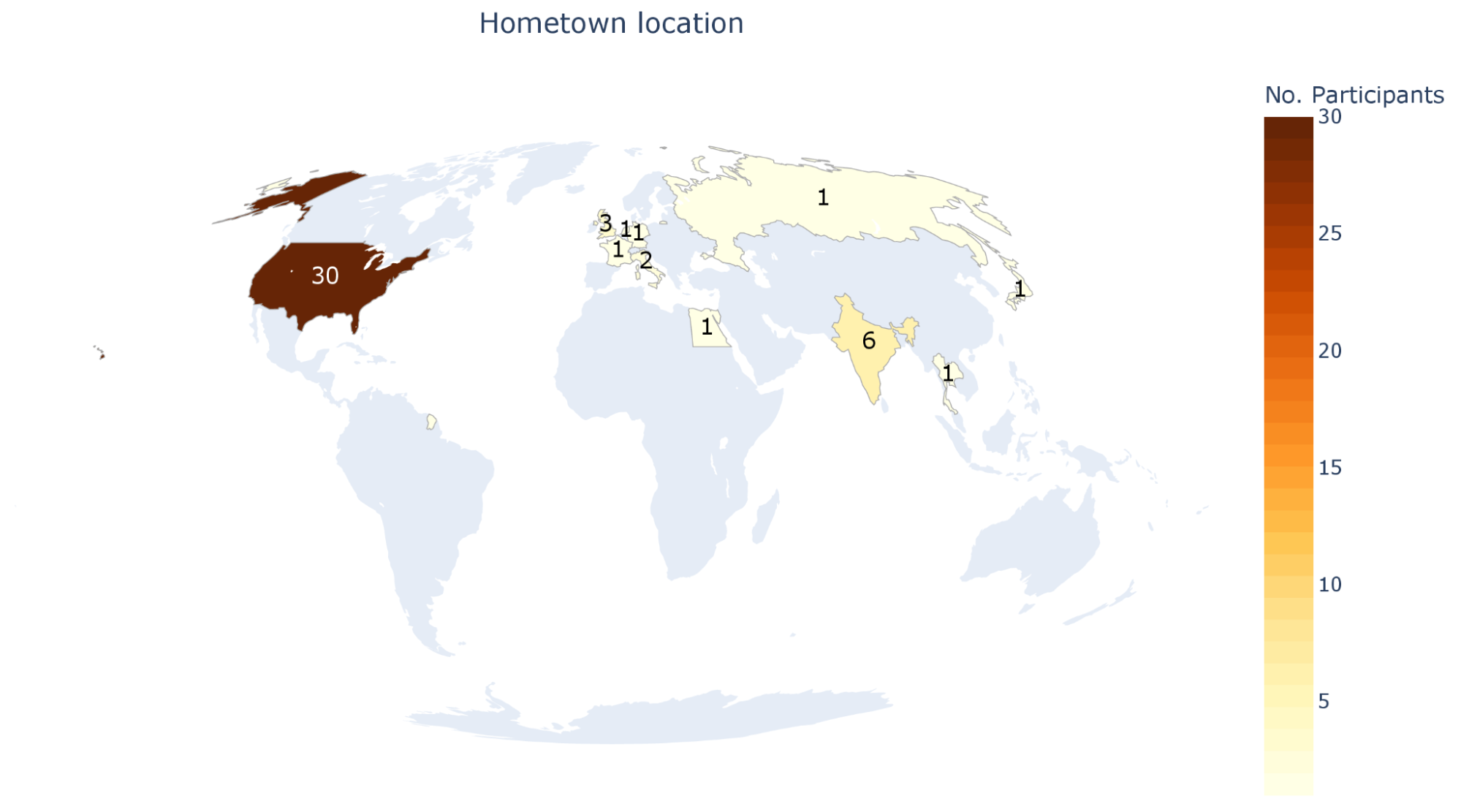}
\caption{\label{fig:HumansofLIGO}
Frequency of Humans of LIGO participants’ institution (top) and hometown (bottom) county. 
}
\end{figure}

Humans of LIGO (\href{https://humansofligo.blogspot.com}{humansofligo.blogspot.com}) showcases the lives, backgrounds, and inspirations of LVK members and GW scientists through blog posts. 
Each post consists of an image of the participant, a direct quote about their experiences, and a short biography including job description and hobbies. 
The blog posts are shared on LIGO social media, and have reached over $53,000$ views (as of May 2024). 
Since its inception in 2018, the blog has featured $59$ participants from $11$ different countries, working at $42$ institutions (Figure~\ref{fig:HumansofLIGO}). 
While profiled scientists come from more diverse locations than just those of LVK institutions, they are still concentrated, reflecting the inequitable distribution of scientific research~\citep{King:2004}. 
A future goal is to add profiles for scientists from other backgrounds and demonstrate potential opportunities within STEM for people across the world.

\begin{figure*}
\centering
\includegraphics[width=.95\textwidth]{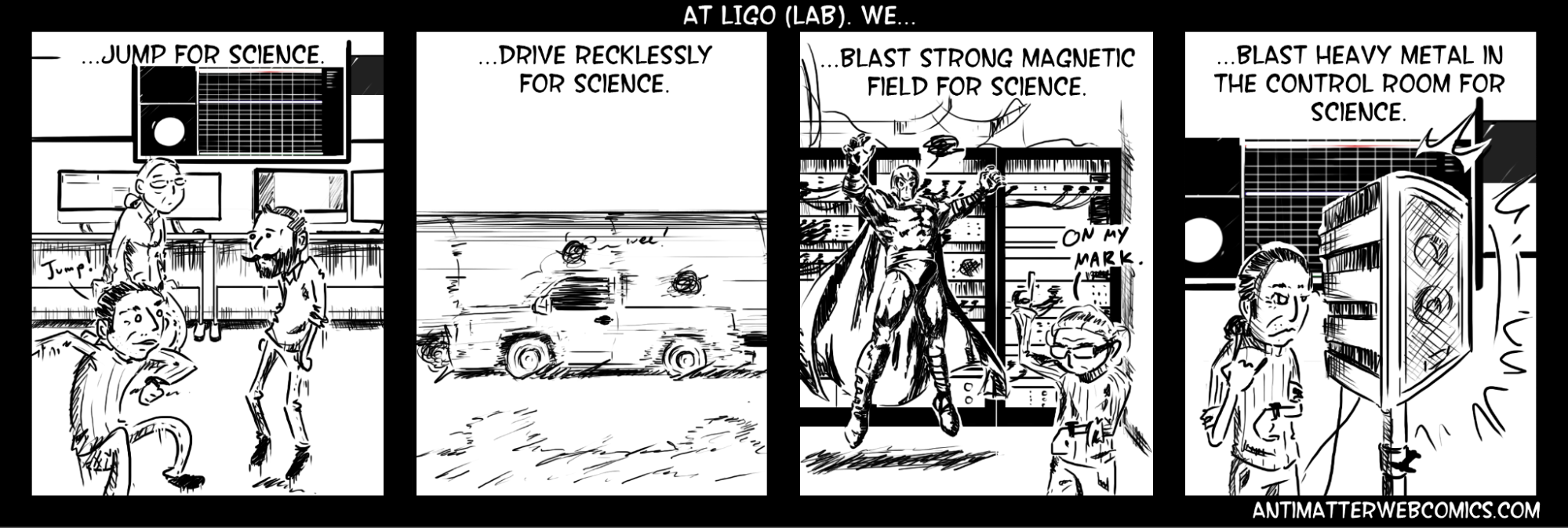}
\caption{\label{fig:Antimatter}
Antimatter comic showing LIGO scientists performing detector checks (credit: Nutsinee Kijbunchoo). 
A version of this comic featured in the \textit{CQG+} blog accompanying a paper describing detector characterization for GW150914~\citep{AbbottEtAlCQGplus:2016}.
}
\end{figure*}

Antimatter web comics (\href{https://antimatterwebcomics.com}{antimatterwebcomics.com}) by Nutsinee Kijbunchoo features day-to-day experiences and current affairs from the GW community and beyond (Figure~\ref{fig:Antimatter}). 
Mental health and graduate-student life are focuses of many comics. 
Based upon webpage views (August 2024), the most popular two comics are about the discovery of the first binary neutron star signal and about living with depression, demonstrating that both communicating discoveries and sharing the lives of the scientists behind them have appeal. 
Comics have been regularly featured in the LIGO Magazine.

The LIGO-India blog, Glorious Women~\citep{LIGOIndiaGloriousWomen} showcases STEM role models, highlighting women from across the LVK. 
Behind-the-scenes interviews feature the daily lives of students, their motivation, contributions and challenges~\citep{LIGOIndiaBehindTheScenes}.

\subsection{Engaging academia}

Sharing our discoveries, and how to use them, is key to ensuring a wider impact on the scientific community. 

Communication with academics is primarily through scholarly articles. 
Since GW astronomy is new, the LVK have written dedicated articles on introductory GW data analysis~\citep{AbbottEtAlGuide:2020}, our data releases~\citep{AbbottEtAlDataReleases:2021, AbbottEtAlOpenDataO2:2023}, and the basic physics of a binary coalescences (suitable for undergraduate teaching)~\citep{AbbottEtAlGW150914BasicPhys:2017}.

Data releases are coordinated by the GW Open Science Center (GWOSC), which hosts raw data, links to LVK publications and further data products (\href{https://gwosc.org/}{gwosc.org}). 
GWOSC maintains lists of open-source analysis software, and provides teaching on data analysis through annual Open Data Workshops. 
Workshops are delivered in a hybrid format, with $200$--$300$ people participating in-person each year, and over $7,700$ people enrolling online over the last four years. 
Workshop materials are openly available online.

To accompany the release of key papers, the LVK organises Zoom webinars, with attendees able to submit questions. 
These provide convenient, direct communication with other scientists, and reach a larger audience than many in-person conferences. 
Recordings are uploaded to YouTube. YouTube views range from $400$ to $3,400$ (August 2024), with the older recordings accumulating more views. 

The LVK also engages with the academic community to promote proper recognition of its three constituent collaborations. 
There is currently a bias in scholarly literature: an abbreviated narrative attributes credit to LIGO, excluding Virgo and KAGRA from LVK achievements, with consequences for scientific careers, accuracy of media and interactions with funding agencies. 
A year-long project~\citep{BarneoEtAl:2024} studied this cognitive bias, and tried to educate the community. 
While intervention has encouraged authors to include proper attribution in their work, it has yet to be seen if this has a long-term impact of reducing incomplete attribution.

\subsection{Supporting formal classroom education}

LVK members have maintained long-term efforts to introduce GWs and related topics into formal classroom education. 
Examples include an Educator's Guide~\citep{Edeon:2016}, courses for community-college teachers, high-school resources, and undergraduate lab demonstrations~\citep{GardnerEtAl:2022}.
 
Einstein-First (\href{https://www.einsteinian physics.com}{www.einsteinian physics.com}) has developed an eight-year spiral curriculum \textit{Eight Steps to Einstein’s Universe} for students aged $7$–$8$ to $15$–$16$ years old. This introduces fundamental concepts (e.g., atoms, molecules, photons and phonons), and in the last year of compulsory science education, a module about GWs, black holes, climate change and Hubble’s law~\citep{PopkovaEtAl:2023}. 
The Einstein-First team is upskilling almost $100$ teachers through micro-credentials courses with many teachers now delivering the program. 
Micro-credential training successfully empowered teachers: $96\%$ agreed or strongly agreed both that they understood why Einsteinian concepts were important to the school curriculum, and that they would feel confident running Einstein-First activities for their students~\citep{KaurEtAlPartA:2023,KaurEtAlPartB:2023}. 
Efforts are underway to introduce Einstein-First in Australia, India, USA, Greece, and Brazil.

Outreach to schools and STEM education programs provides an avenue for sharing GW discoveries, engaging students and teachers, and additionally provides science-communication training for GW researchers. 
Assessment of the impact of the Space Public Outreach Team (SPOT) program shows positive outcomes for students at all levels for SPOT programs across a range of geographic locations, benefitting the college-student presenters with improved scientific knowledge and presentation skills, and the audience of school children with greater engagement with science~\citep{KeyEtAl:2024,DesJardinsEtAl:2020}.

\subsection{Writing reference texts}

\begin{figure*}
\begin{center}
\includegraphics[width=0.8\textwidth]{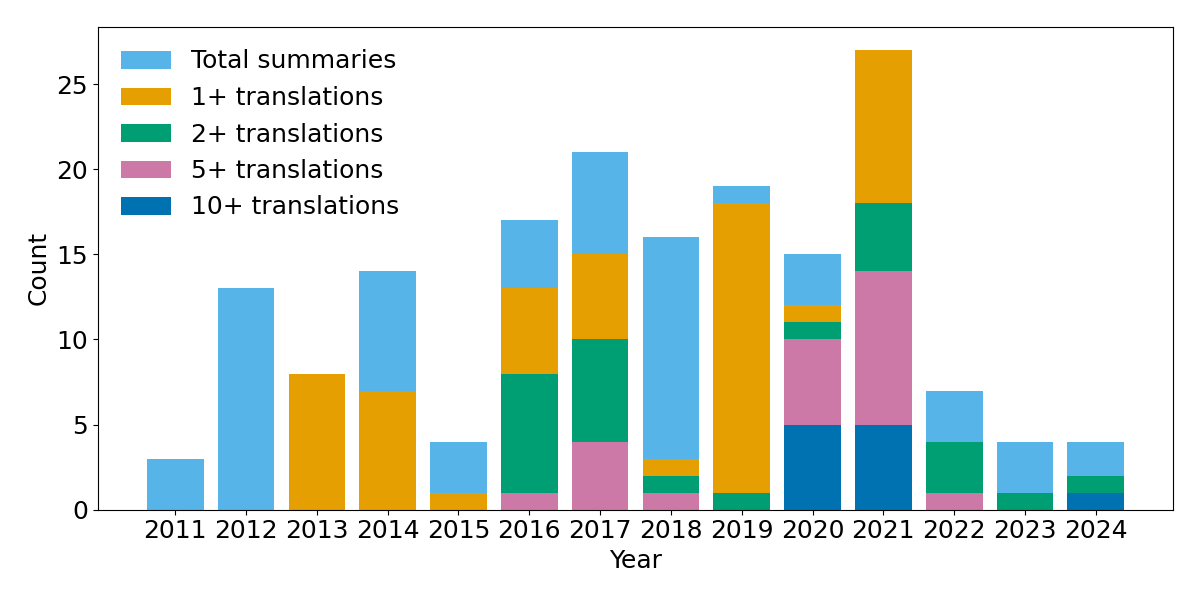}
\caption{\label{fig:Summaries}
Number of Science Summaries by year of posting, and their number of translations. 
There were fewer papers in 2021–2024 while waiting for the results of the latest observing run.
}
\end{center}
\end{figure*}

LVK Science Summaries and press releases are both Collaboration-coordinated activities to provide written summaries of our discoveries.

LVK Science Summaries are a long-standing effort to explain LVK publications at a technical level typically higher than in the popular press but still accessible for enthusiastic lay readers. 
They are published on \href{https://www.ligo.org}{www.ligo.org}, distributed as print copies at outreach events, and shared with journalists. 
Summaries for discovery announcements may get thousands to tens of thousands of views around publication. 
Summaries are translated into a variety of languages by LVK volunteers, with discovery papers often receiving the most attention (e.g., the GW190521 summary was translated into 16 languages)~\citep{KeitelEtAl:2021}. 
Numbers of summaries and translations over time are displayed in Figure~\ref{fig:Summaries}, demonstrating significant growth in translation activity especially during the LVK’s third observing run (2019–2020). 
However, translation activity is typically less diverse for summaries of lower-profile (non-discovery) papers, demonstrating a similar pattern of enthusiasm by LVK translators and the target audience

Press releases to accompany discovery announcements are prepared centrally, including quotations from a selection of LVK members, and then adapted by member institutions (often including quotations from their scientists). 
As a global collaboration, it is not possible to find a single time for press releases to be made public that works for all time zones, which can make it challenging to coordinate local coverage.

Both Science Summaries and press releases have been drawn upon by journalists~\citep[e.g.,][]{Carlise:2021}. 
Translation of Science Summaries is especially important for use in non-English media. 

\subsection{Interacting in person}

In-person outreach to the general public reaches fewer people than other means, but it can yield impactful interactions, and may be perceived as having more value~\citep{Baucum:2022}. 
As reported by one young attendee to in-person events: ``one of the activities that most fostered my connection to the STEM world was attending in-person talks, conferences, and panel discussions […] ultimately leading me to pursue a degree in physics.''

LVK members participate in science festivals, classroom visits, the creation of museum exhibits (such as interactive detector models~\citep{CooperEtAl:2021}, or the \textit{Black Hole and Gravitational Wave} circular exhibitions, lectures and panel discussions across Taiwan and Japan), visitor centers, and tours of the GW observatories. The face-to-face format allows staff to customize the experience to the visitor.
 
LVK’s visitor centers merge artifacts from the detectors with interactive exhibits that put the visitor in the role of a scientist or engineer. 
The exhibits explore the underlying science and engineering of GW detectors, from the importance of pendula to concepts underpinning gravity. 
In-person and virtual tours of the detectors give a glimpse to the heart of operations. 
Over $330,000$ visitors have interacted with the LVK visitor centers; 2024 attendance is projected to reach $30,000$. 
 
Science festivals and local outreach efforts around the world connect with classrooms and the general public. 
Outreach is performed by seasoned scientists, engineers, graduate students and undergraduates new to GW science. 
Such efforts positively impact the undergraduates who perform the outreach~\citep{Carpenter:2015, DocentSurvey:2014, YoungKatzman:2023}. 
A 2023 survey revealed that $95\%$ of undergraduates said ``their experience has changed the way they interact with their communities'' and over half of LIGO Livingston program undergraduates said it ``helped them decide to go into a STEM field or educational field''~\citep{DocentSurvey:2023}. 
Undergraduates often note soft-skills development, one said ``My experience as a LIGO docent has changed my life and how I interact with everyone. From conducting my presentations at school, teaching my kiddos, conducting business, or simply talking to others.''

\subsection{Creating graphics}

\begin{figure*}
\centering
\includegraphics[width=0.75\textwidth]{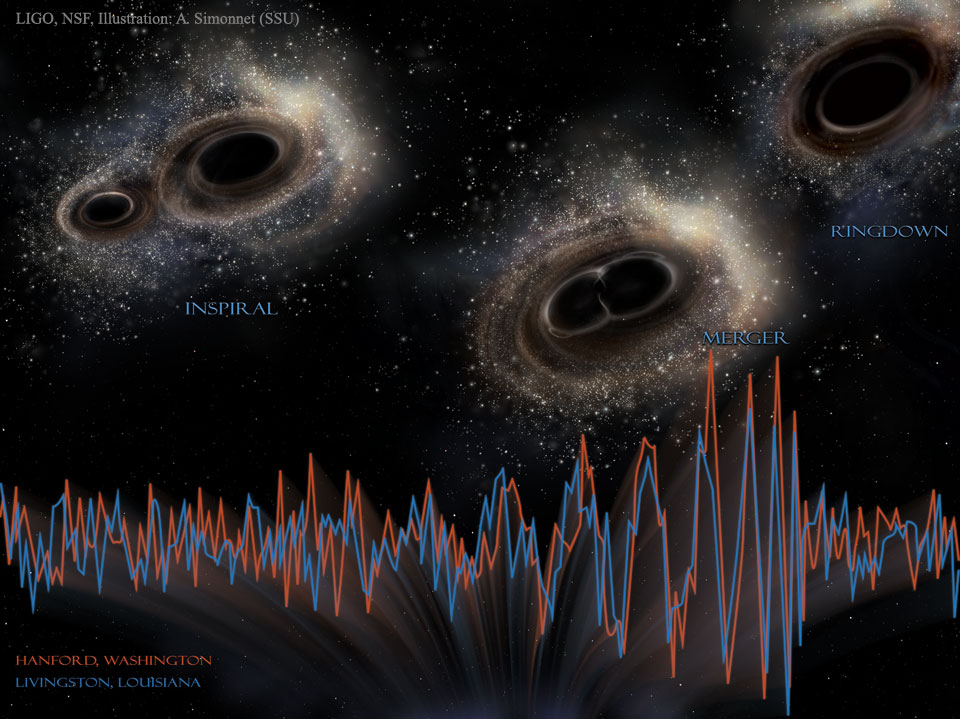}
\caption{\label{fig:GW150914}
The detector data for GW150914, and an artistic representation of its binary black hole source (credit: LIGO/National Science Foundation/Aurore Simmonet/Somona State University). 
}
\end{figure*}

Unlike electromagnetic observatories, GW data does not naturally lend itself to the stunning images that are commonplace throughout astronomy. Nevertheless, visualisations, infographics and cartoons~\citep[e.g.,][]{KijbunchooEtAl:2016, SpectraThompsonEtAl:2020} form key parts of our outreach activities, e.g., infographics displaying key facts accompany discoveries; simplified plots of results may be created for Science Summaries, and artistic representations of our sources may accompany press releases

Videos and images are also produced from numerical-relativity simulations of sources and the GWs they emit. 
Since new simulations are often produced to better understand the physics of novel sources, fresh visualisations have accompanied discovery announcements. 

Images accompanying LVK discovery announcements have been selected as NASA Astronomy Pictures of the Day, e.g., to accompany the first detection, an artistic representation of a binary black hole coalescence (Figure~\ref{fig:GW150914}) and a corresponding numerical-relativity simulation were featured (11 and 12 February 2016), and to accompany GW190521, an artistic representation of the GW emission from two spinning black holes was featured (8 September 2020).

\subsection{Employing multisensory resources}

Sound is a common analogy when describing GWs. 
While GWs are not sound, the GW frequencies observable by the LVK are similar to audio frequencies of human hearing. 
The signals lend themselves to audification (\href{https://www.soundsofspacetime.org}{www.soundsofspacetime.org}): the technical name for the signal from a binary inspiral is a ``chirp'' in reference to its sweep up in frequency.

GWs can also be communicated through touch. 
The GW150914 signal has been translated to a 3D-printable model, specifically tailored to the needs of visually impaired people. 
By running fingers along the edges of the shape, one can appreciate the features of a GW and perceive the details of the signal evolving in time. 
This design is freely available~\citep{EGOVirgoTactileGW150914:2021}, along with suggestions for classroom activities. 

\begin{figure}
\centering
\includegraphics[width=0.48\textwidth]{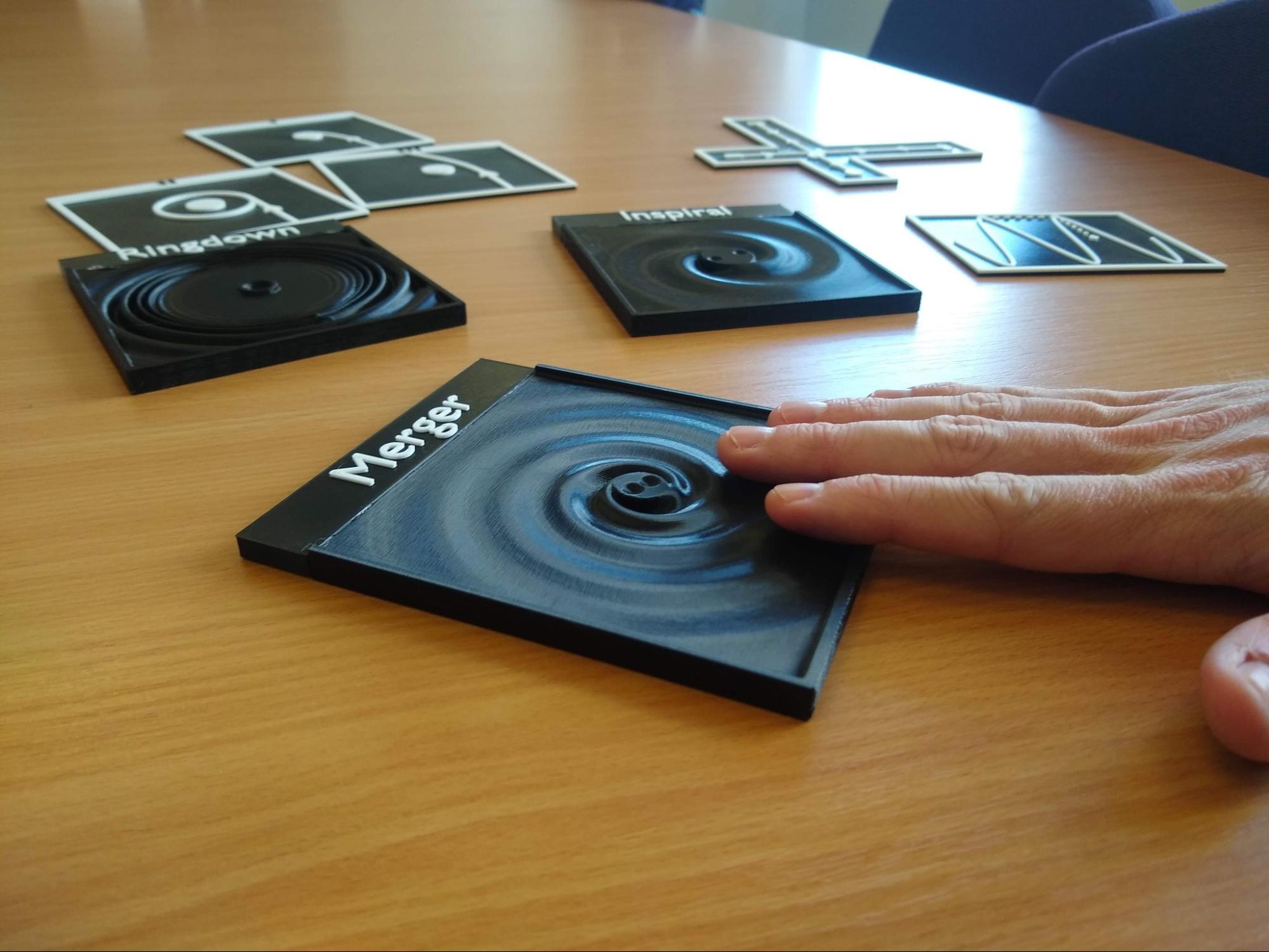}
\caption{\label{fig:tactileUniverse}
A Tactile Universe 3D printed surface of the inspiral of two black holes and the resulting GWs 
}
\end{figure}

The Tactile Universe (\href{https://www.tactileuniverse.org}{www.tactileuniverse.org})~\citep{BonneEtAl:2018} has partnered with LVK groups to design tactile resource sets for use with blind and vision-impaired pupils in upper secondary education. A mix of 3D-printable (Figure~\ref{fig:tactileUniverse}) and more basic tactile resources, paired with existing sonifications, are tied to $20$~minute workshops about black holes and neutron stars, GW signals and detection, and core science topics like waves and gravity. 
The 3D-printable resources, example lesson scripts and guides are available on the website. 
Students have enjoyed interacting with the tactile resources; teachers have praised the accessibility and pacing of the workshops, and in post-session feedback, students have demonstrated an understanding of the concepts discussed.

The audifications and tactile resources have also proved effective at science festivals, with members of the public appreciating having the diverse means of explaining unfamiliar concepts. 
This demonstrates how providing multisensory resources makes ideas accessible to a wider audience as well as more comprehensible to existing audiences.

\subsection{Combining art and music with science}

Art, music, and science collaborations explore novel approaches to GW communication, showcase the science through a different point of view, and bring the topic to new audiences. 
Works have been created by professional artists and musicians, and LVK members~\citep[e.g.,][]{AzureEtAl:2021}.
Examples include music compositions~\citep{InstitutodeFisicaCorpuscular:2023, PenguinCafe:2023}, open land art installations~\citep{VirgoLandArt:2022}, museum exhibits~\citep{EGORythmOfSpace:2019} and novel musical instruments such as GravitySynth which combines the technology of GW detection with modular synthesisers~\citep{Trimble:2024, AzureEtAl:2021}. 
The LIGO-India blog has a virtual gallery GWSciArt that showcases GW-inspired artistic projects~\citep{LIGOIndiaGWSciArt}.

Combining art with science can help reach new audiences and provide a memorable experience. For example, the art and science festival \textit{Celebrating Einstein} merged dance, music, and film with GWs. 
Assessment of the festival audience demonstrated that those who attended were a mix of those interested in science and art; that non-physics experts gained knowledge (e.g., $75\%$ of physics novices improved their score between pre- and post-event surveys), and that attendees typically had a positive emotional response to the event~\citep{GrimbergEtAl:2019}. 
Similarly, \textit{Into the Quadrivium}, a collaboration between GW researchers and musicians, which blended baroque and contemporary music with spoken-word explanations of GW science, was positively received by the audience~\citep{IntoTheQuadrivium:2023}.

\subsection{Communicating through social and non-traditional media}

Non-traditional media provides an opportunity to present our science to broad audiences. 
Activities like social-media posts, blogs, and podcasts or interviews can communicate science in an informal way. 

Interviews have been hosted on the Spanish podcasts ``Oscilador armónico'' ~\citep{CoderoCarrionEtAl:2023} ($7,800$ listens as of August 2024), ``A ciencia cierta''~\citep{Rivera:2021} ($33,100$ listens) and ``Coffee Break: Señal y Ruido''~\citep{SocasNavarro:2021} ($40,800$ listens). 
The last podcast coined the term ``gravitondas'' (``gravi-waves'') for GWs. 
Communicating with audiences in their local language makes science more accessible. 
The podcast format is more flexible than traditional radio programming allowing more time and in-depth discussion needed to explain complicated ideas.

The LVK has several social-media accounts, for the Collaborations and the observatory sites, across different platforms (Table~\ref{tab:socialMediaFollowers}). 
The greater following for LIGO compared to Virgo and KAGRA may be a consequence of the same bias that leads to LIGO preferentially receiving credit. 
Content includes educational resources, news stories, and discovery announcements.
Special posts are scheduled for events such as detection anniversaries or the International Day of Women and Girls in Science. 
The observatory accounts often share posts about local news, such as pictures from the sites or events at the science centres, while the Collaboration accounts take the lead on big announcements. 

\begin{table}
\begin{center}
\caption{\label{tab:socialMediaFollowers}
Social-media followers (August 2024).
}
\begin{tabular}{p{0.085\textwidth} p{0.26\textwidth} r}
\toprule
\textbf{Platform} & \textbf{Account} & \multicolumn{1}{l}{ \textbf{Followers}} \\
\midrule
X  
  & LIGO Scientific Collaboration
  & $110,600$ \\ 
  & European Gravitational Observatory and Virgo 
  & $12,500$ \\
  & KAGRA 
  & $2,500$ \\
  & LIGO Hanford Observatory
  & $10,800$ \\
  & LIGO Livingston Observatory
  & $5,300$ \\
  & LIGO India
  & $6,700$ \\
  & LIGO Magazine
  & $690$ \\ 

Facebook
  & LIGO Scientific Collaboration
  & $33,000$ \\
  & European Gravitational Observatory and Virgo
  & $7,000$ \\
  & LIGO Hanford Observatory
  & $6,100$ \\
  & LIGO Livingston Observatory
  & $6,800$ \\
  & LIGO India
  & $9,600$ \\

Instagram
  & LIGO–Virgo
  & $14,900$ \\
  & LIGO India
  & $5,600$ \\

Mastodon (Astrodon)
  & LIGO Scientific Collaboration
  & $1,500$ \\
\bottomrule
\end{tabular}\textbf{}
\end{center}
\end{table}

Social-media content is reshared across platforms with suitable adaptations. 
The microblogging X and Mastodon platforms are well-suited to threads; these are useful for linking many resources or explaining a topic from many angles. 
Facebook and Instagram allow for longer posts, enabling more in-depth explanation per post, but posts must be spaced further apart to avoid cluttering followers' feeds. 
Across all platforms, the most popular posts often feature a graphic or video (all Instagram posts include a graphic or video), and hence these are sought whenever possible.

The different social-media platforms reach different numbers of people and demographics. 
Table~\ref{tab:socialMediaComparison} shows metrics for the GW230529\_181500 discovery announcement (April 2024). 
While X has the most views/users reached; Mastodon has the most shares per follower (as a newer platform it potentially has a higher proportion of active users), and Instagram has the most likes per follower.

\begin{table*}
\begin{center}
\caption{\label{tab:socialMediaComparison}
Social-media performance as of August 2024 for LIGO posts about GW230529\_181500 (GW230529) and the GWTC-3 Orrery. 
Statistics are not directly comparable between platforms but qualitatively similar. 
Numbers per follower are given in parentheses.
}
\begin{tabular}{p{0.1\textwidth} p{0.22\textwidth} r l r l r l r l }
\toprule
     &        & \multicolumn{8}{c}{\textbf{Platform}} \\ \cline{3-10}
\textbf{Post} & \textbf{Metric} & \multicolumn{2}{l}{\textbf{X}} & \multicolumn{2}{l}{\textbf{Facebook}} & \multicolumn{2}{l}{\textbf{Instagram}} & \multicolumn{2}{l}{\textbf{Mastodon}} \\ 
\midrule
GW230529 
         & Reposts/shares/boosts
         & $155$ & ($0.0014$)
         & $89$  & ($0.0027$)
         & $72$  & ($0.0048$)
         & $18$  & ($0.0120$) \\
         & Likes/reactions/favorites
         & $328$ & ($0.0030$)
         & $261$ & ($0.0079$)
         & $662$ & ($0.0444$)
         & $26$  & ($0.0173$) \\
         & Views/reach
         & $46,000$ & ($0.42$)
         & $18,300$ & ($0.55$)
         & $5,200$  & ($0.35$)
         & \multicolumn{2}{l}{n/a}\\

Orrery
         & Reposts/shares/boosts
         & $113$ & ($0.0010$)
         & $48$  & ($0.0014$)
         & $51$  & ($0.0034$)
         & $25$  & ($0.0167$) \\
         & Likes/reactions/favorites
         & $350$ & ($0.0032$)
         & $144$ & ($0.0043$)
         & $813$ & ($0.0546$)
         & $32$  & ($0.0213$) \\
         & Views/reach
         & $192,000$ & ($1.74$)
         & $7,800$   & ($0.24$)
         & $9,600$   & ($0.64$)
         & \multicolumn{2}{l}{n/a} \\
\bottomrule
\end{tabular}
\end{center}
\end{table*}

Social media allows the LVK to interact directly with the public, and threads in response to questions have often proved popular. 
On Reddit, the LVK has run occasional Ask-me-anything discussions. 
The Ask-me-anything accompanying GW150914 has $2,300$ comments. 
Posting answers to questions online enables them to be searched out in the future.

\subsection{Using interactive technologies}

Interactive apps and games are engaging and fun ways to convey GW science. 
Interactive apps and games are engaging and fun ways to convey GW science. 
Physical interaction with virtual environments incorporates additional modes for audiences to engage with content. 
Activities have ranged from computer games~\citep{CarboneEtAl:2012} (\href{https://www.laserlabs.org/}{www.laserlabs.org}; \href{https://blackholehunter.org}{blackholehunter.org}) to classroom and science festival teaching with virtual reality (\href{https://www.scivr.com.au/}{www.scivr.com.au})~\citep{Parks:2023} .

Mission Gravity~\citep{BondellandMyers:2021,KerstingEtAl:2024} is an interactive virtual-reality environment designed for secondary school students to collaborate to better understand black holes and neutron stars. 
Since 2018, the program has been delivered to over 23,000 students across over $400$ schools, as well as to over 1,100 teachers through professional-development workshops. 
One student reported: ``The virtual reality component of it was engaging and enjoyable […] I was able to understand ideas that I previously struggled with.''

\section{Case Study: GWTC-3}

GW discoveries from the LVK are published in catalogs: papers accompanied by data releases. 
GWTC-3~\citep{AbbottEtAlGWTC3:2023} is the most recent of these, presenting results up to the end of the third observing run. 
Thanks to the continued improvements in detector sensitivity~\citep{AbbottEtAlObsScenario:2020}, the rate of discoveries has increased with time, meaning that each catalog includes a significant number of new GW candidates. 
GWTC-3 represents the most sophisticated and comprehensive analysis published by the LVK to date.

Production and publication of a catalog is a multi-year project directly involving hundreds of experts. 
Given the significance of the results and the importance of easy-to-use data releases, effective communication is a priority. 
However, communication of catalog results faces challenges: catalog results consist of comprehensive analyses of many signals (making them overwhelming), and the majority of detections have properties similar to those observed in the past (lacking the excitement of a novel discovery). 

For GWTC-3, the project team included a member responsible for coordinating communication and outreach activities from the beginning. 
This embedding allowed for these activities to be planned as the project progressed, developed in tandem with the data releases, and coordinated with those compiling the results. 
Delaying this work until the project was nearly complete may not have left sufficient time to produce and review resources, and risked the team being too exhausted to contribute to these tasks. 
LVK paper teams typically have a designated person responsible for producing data products and for writing the Science Summary, but these may not necessarily be integrated into the team throughout the project, and there is not typically one person responsible for coordinating communication and outreach more generally. 
We recommend following an approach similar to GWTC-3 of having a dedicated team member responsible for communication and outreach, and starting work on the resources early. 

The variety of activities designed to communicate the discoveries from GWTC-3 span the diverse range of audiences targeted and platforms used by the LVK.

\subsection{Sharing the people behind the science}
LIGO Magazine issue $20$ featured an $8$ page article by $23$ members of the analyses, paper-writing, and engagement teams~\citep{BerryEtAl:2020}. 
They wrote about their experiences of working on the catalog and associated results. 
Although the article authors represent only a fraction of those providing input to GWTC-3, features like this give readers a behind-the-scenes insight of working on big LVK results. 

\subsection{Engaging academia}
A series of LVK webinars were organised to coincide with the release of GWTC-3. 
The first on GWTC-2.1~\citep{AbbottEtAlGWTC2p1:2024}, a reanalysis of previous data using updated methods consistent with GWTC-3. 
Subsequent webinars presented the GWTC-3 results and implications for cosmology, astrophysics and tests of general relativity. 
Presenters were drawn from those who made key contributions to the work, with preference given to early-career scientists, and balancing geographic location and demographics.

The data release was a key component of the GWTC-3 results. 
It included Jupyter notebooks and Python scripts to demonstrate use and how to reproduce plots from the paper. 
As of August 2024, the data release for the inferred source properties of the new GWTC-3 detections is the most viewed of all Zenodo-hosted LVK data releases, with approximately $8,000$ views ($27,500$ file downloads); the second most viewed is the equivalent set of results for the GWTC-2.1 analysis, with approximately $5,600$ views ($21,800$ file downloads). 
The popularity of the data release indicates that the community appreciates the value of the results, and understands how to use data products.

As the detection rate increases, it becomes more difficult to manage all the analyses, and curate data releases that include all the relevant results (especially if there have been multiple reanalyses to obtain final results). 
We recommend automating the process to reduce the chance of human error in data-release production, and efforts are underway to develop such automation tools~\citep[e.g.,][]{WilliamsEtAl:2023}.

For GWTC-3, a Streamlit app was created to interactively make plots of source parameters enabling users to customise paper plots without needing to download data (\href{https://gwtc3-contours.streamlit.app/}{gwtc3-contours.streamlit.app/}).

\subsection{Writing reference texts}
Like the paper itself, the GWTC-3 Science Summary~\citep{GWTC3ScienceSummary:2021} had a broad scope to cover.
It was written in close collaboration between outreach experts and the analysis and paper-writing teams. 
It includes background information needed to understand the catalog, and a subset of highlight discoveries. 
The summary has been translated into $9$ languages.

A LVK news item, rather than a press release, was drafted for GWTC-3. 
This was used by individual institutions to draft their press releases. 
The decision to not have a LVK press release reflected the expectation that GWTC-3 would be more difficult to communicate, and hence less attractive to write about, than a single discovery. 
Despite this, several science journalists did write articles about GWTC-3~\citep[e.g.,][]{Carlise:2021, Castelvecchi:2021, Plait:2021}. 
This demonstrates that the LVK did at least partially succeed in efforts to communicate the science of GWTC-3 to journalists, and (assuming similar communications efforts are organised in the future) that there may be an audience for press releases for future catalog releases.

Resources, including the news item and institutional press releases, were made public at 01:00 UTC on Monday 8 November 2023. 
The time was chosen to coincide with the paper preprint appearing on arXiv, while the date was chosen to correspond to the predefined detector data release timeline. 
An embargo period was offered from 18:00 UTC on Thursday 4 November when trusted journalists could be contacted about the upcoming release. 
Providing such an embargo period gives time for journalists to prepare and check their stories, which is beneficial; however, this embargo period was mostly over the weekend. 
Therefore, for future discovery announcements, it may be beneficial to have a longer embargo period, or to choose the day of the week for release more carefully. 

\subsection{Creating graphics}

\begin{figure*}
\centering
\includegraphics[width=0.7\textwidth]{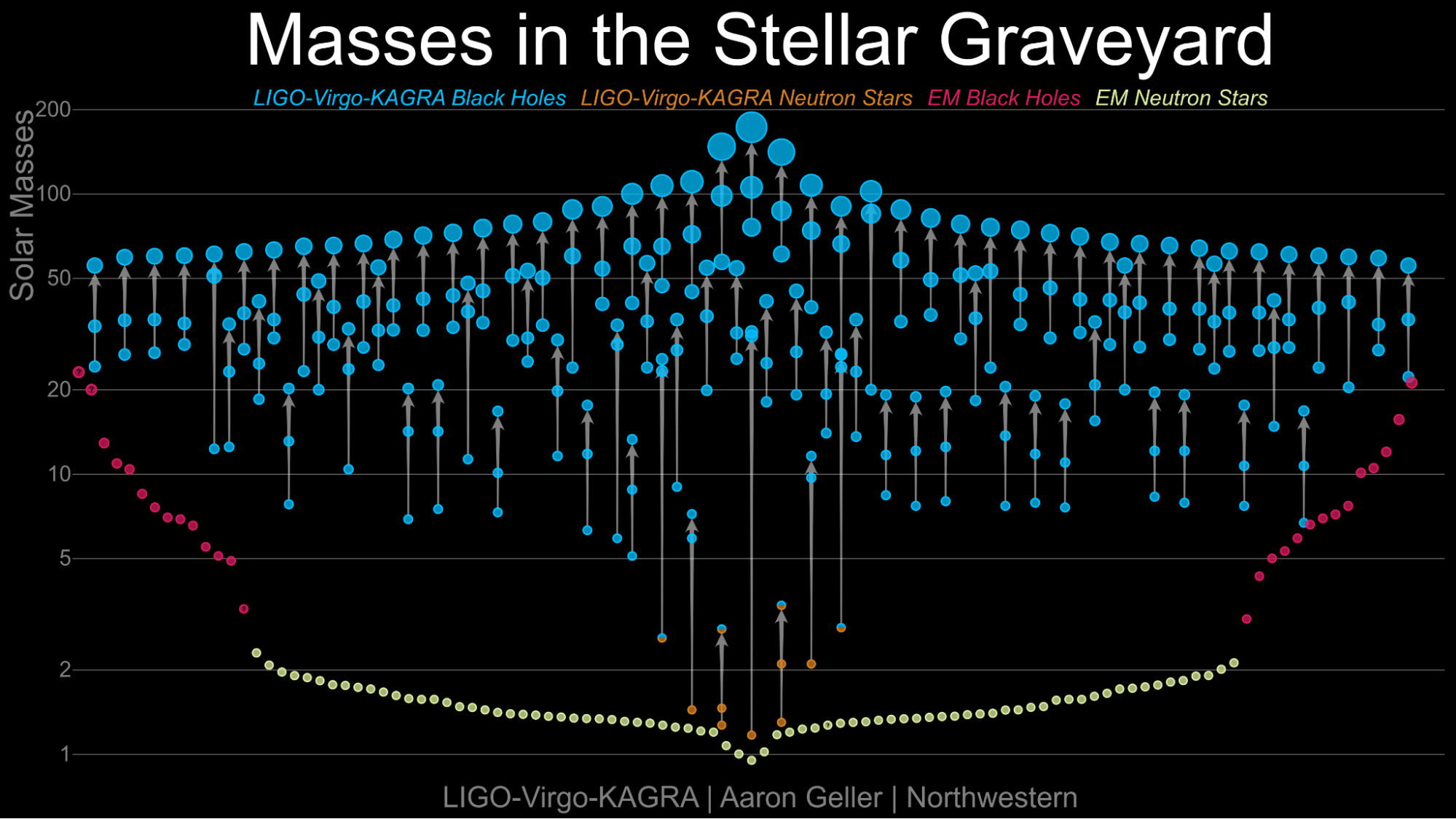}
\caption{\label{fig:GWTC3Masses}
GWTC-3 version of Masses in the Stellar Graveyard (credit: LVK/Aaron Geller/Northwestern). 
The plot shows the known masses of stellar-mass black holes and neutron stars observed electromagnetically and with GWs.
}
\end{figure*}

Updates to some long-standing LVK visualisations were made for GWTC-3. 
The new results were added to the Masses in the Stellar Graveyard plot (Figure~\ref{fig:GWTC3Masses}). 
This plot is now generated from GWOSC data products, making it easy to add large numbers of detections. 
An online version allows for customisable images~\citep{GellarEtAl:2024}. The LVK Orrery was also updated, which shows a stylised animation of the GWTC-3 binaries. 

\begin{figure}
\centering
\includegraphics[width=0.45\textwidth]{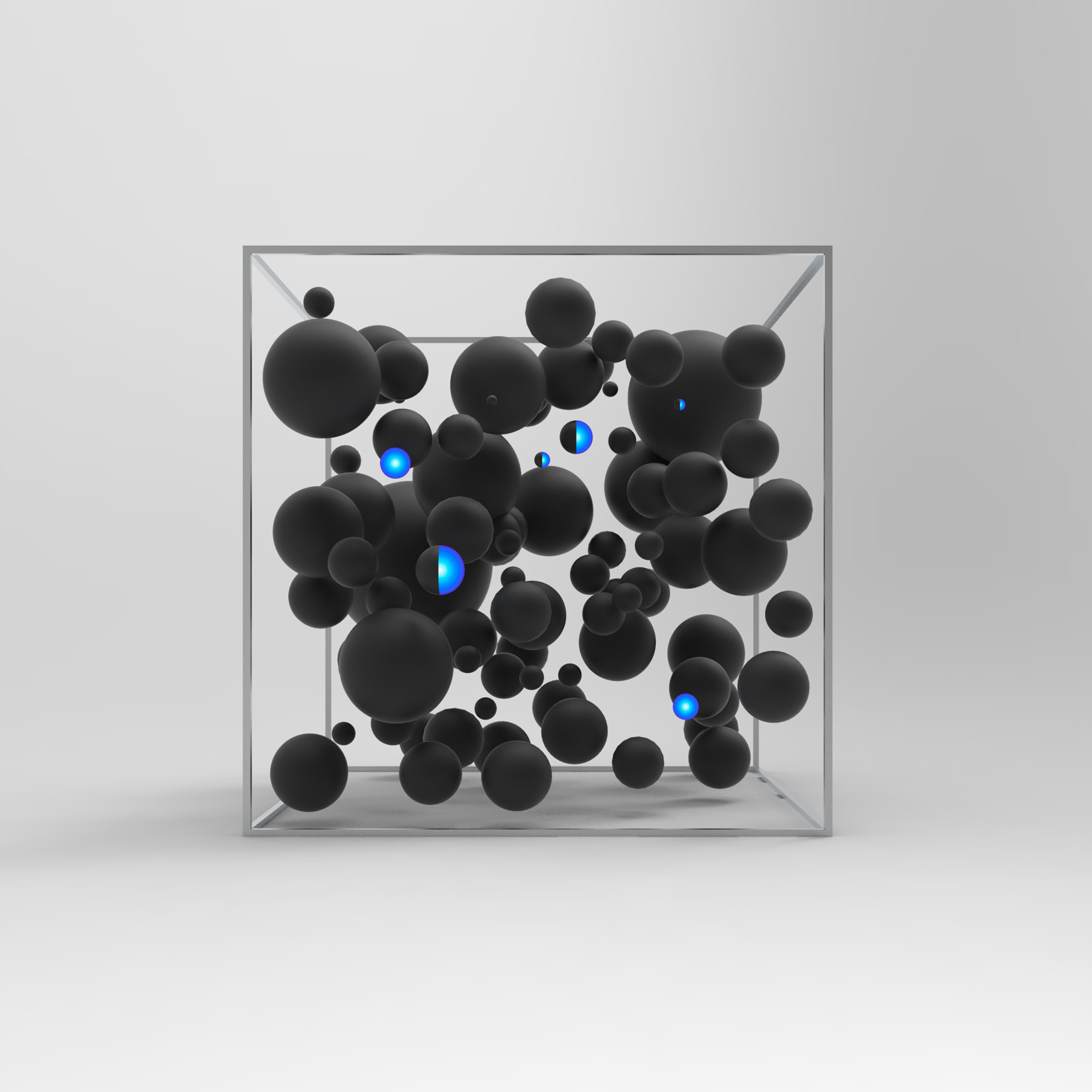}
\caption{\label{fig:GWTC3InABox}
An artistic representation of GWTC-3 sources which accompanied the journal publication (credit: Carl Knox/OzGrav Swinburne). 
}
\end{figure}

New visualisations were also created. 
One was a poster reminiscent of a periodic table showing the catalog after each observing period. 
The poster was used in a variety of articles, a double-page LIGO Magazine spread~\citep{Knox:2022}, and in the GWTC-3 webinar. 
An artistic representation of the detected sources in a box accompanied the publication in \textit{Physical Review X} (Figure~\ref{fig:GWTC3InABox}). 
An image showing spectrograms for the $90$ GWTC-3 signals was featured as NASA Astronomy Picture of the Day (7 December 2021).

\subsection{Employing multisensory resources}
Low Mass Beats is a audification of the catalog, converting masses into pitch for each black hole or neutron star: the greater the mass, the lower the pitch. 
This lighthearted project provides an entertaining perspective on the catalog. 
(SoundCloud: \href{https://soundcloud.com/user-889003031}{soundcloud.com/user-889003031})

\subsection{Communicating through social and non-traditional media}
GWTC-3 material was shared via a social-media campaign in two bursts using the hashtag \#GWTC3. 
The first burst coincided with the release of the preprint and associated data. 
Special attention was given to advertising the webinars and data releases. 
The second burst coincided with the publication of the journal article and priority was given to any resources developed since the initial release, such as the Streamlit plotter and Low Mass Beats. 

For both bursts, posts were spaced over multiple days to catch the attention of people in different timezones. 
Material was translated into regional languages~\citep{LIGOIndiapoLIGlOt}.  
The pacing also made it natural to integrate popular-science articles written immediately after the release.

From the second burst, the most popular post was for the Orrery. 
Statistics are shown in Table~\ref{tab:socialMediaComparison}. 
It is generally less shared than the GW230529\_181500 post, but there were more posts for GWTC-3 than for GW230529\_181500 spreading out attention: that numbers are comparable demonstrates that catalog results can capture interest similarly to new detections. 
The Orrery has high X views having been reposted by accounts with large followings. 

Resources were subsequently reused, e.g., Low Mass Beats was shared on Halloween because of its spooky sound, and the Orrery was shared during NASA’s Black Hole Week.

The impact of the social-media campaign can be partially assessed via the Altmetric score. As of August 2024, the GWTC-3 paper has a score of $714$ (including $56$ news outlets and $6$ Wikipedia pages). This score is in the $99\rm{th}$ percentile of all tracked outputs.

\section{Conclusions}
GW astronomy is a new branch of astronomy. 
After decades of research, the first observation was made in 2015, and the field has grown rapidly since. 
These discoveries have been led by the LVK, an international collaboration of thousands of scientists. 
Sharing these discoveries has been a priority for the LVK. 
As a new field, communication has been essential to both publicise and explain results.

LVK activities span a range of audiences and media. 
Within the LVK, different activities are run by different teams, some coordinated Collaboration-wide, while others reflect individual interests. 
This provides a rich environment, where science-communication resources can be shared between different communications efforts. 
Many resources are used in multiple contexts, such that utility can be amplified beyond the original target, e.g., paper plots used on social media, infographics in talks, and data audifications in musical performances.

Communicating LVK discoveries has a unique set of challenges, but each provides an opportunity:
\begin{itemize}
\item GWs are an unfamiliar means of performing astronomy. 
Unlike traditional astronomy, GW astronomy does not directly image sources. 
However, GW data is suitable for audification, and a variety of tactile resources have been made. 
These make GW astronomy accessible to people with vision impairments, who are often excluded from accessible astronomy communication~\citep{BellSilverman:2019}, in addition to providing a novel way to explain these physical concepts. 
\item LVK results are produced by large teams. 
This makes it difficult for those outside to understand how discoveries are made, or to construct human-interest stories. 
However, this enables emphasization of the importance of teamwork and international cooperation to science. 
The LVK’s breakthroughs are a contrast to the popular perception that scientific breakthroughs come from lone geniuses~\citep{LariviereEtAl:2015, AksnesAagaard:2021}. 
The diversity of the LVK’s membership also allows us to draw upon many different lived experiences, and demonstrate that scientists may come from any background.
\item After the initial first observations, many detections were similar to those discovered previously. 
The increase in the LVK’s detection rate is an experimental triumph---what was once extraordinary is now everyday. 
It is therefore necessary to concentrate on the science enabled by a large set of detections. 
This has a benefit of reflecting that most scientific progress comes from careful study rather than a dramatic breakthrough. 
\end{itemize}
Elements of each challenge may translate to other communication efforts: 
\begin{itemize}
\item When abstract or theoretical concepts are being communicated, having diverse explanations (combining multiple views of the concept and engaging multiple senses) may aid audience comprehension. 
\item When results come from large collaborations, which is increasingly common in physics and astronomy~\citep{Smith:2016,BattistonEtAl:2019}, drawing upon the diverse experiences of collaboration members may help engage members of the public with similar backgrounds. 
\item When a new field is founded, charting the way the field will evolve long term may help increase scientific literacy regarding how scientific progress is made through many small advancements (in addition to through big discoveries).
\end{itemize}
Consequently, while we do not expect the exact circumstances faced by the LVK will be replicated, we anticipate that insights into GW-science communication may benefit others across science.

In communicating GW discoveries, the LVK has benefitted from existing interest in black holes and space. 
It has also faced hurdles, such as a tendency to misattribute discoveries to LIGO alone. 
This highlights the importance of taking a holistic approach to planning and reviewing communication: it is advantageous that an audience correctly understands how science is done, as well as what the discoveries are, and it cannot be assumed that just because the audience is interested they will absorb all relevant information.

As GW astronomy continues to mature, we expect that communication strategies will need to evolve to reflect both the state of the field, and audience interest and awareness.

\section*{Acknowledgements}
We thank Jess McIver, the listeners of Coffee Break: Señal y Ruido, and the anonymous referees for constructive feedback on this manuscript. 
This material is based upon work supported by LIGO Laboratory which is a major facility fully funded by the US National Science Foundation and by the European Gravitational Observatory (EGO) which is funded by Centre National de la Recherche Scientifique (CNRS) in France, the Istituto Nazionale di Fisica Nucleare (INFN) in Italy, and the National Institute for Subatomic Physics (Nikhef) in the Netherlands 
This paper has been assigned LIGO DCC number P2400039 (\href{https://dcc.ligo.org/P2400039/public}{dcc.ligo.org/P2400039/public}). 

The Tactile Universe project was jointly funded by the University of Portsmouth and an STFC Public Engagement Legacy Award [ST/V001515/1]. 
HM is supported by the UK Space Agency Grant Nos.\ ST/Y004922/1; ST/V002813/1; ST/X002071/1. 
CPLB acknowledges support of STFC grant ST/V005634/1. 
IC-C acknowledges support by the Spanish Agencia Estatal de Investigación through the Grant No.\ PID2021-125485NB-C21 funded by the MCIN/AEI/10.13039/501100011033, by the Generalitat Valenciana (Prometeo program for excellent research groups) through the Grant No.\ CIPROM/2022/49 and the Astrophysics and High Energy Physics program Grant No.\ ASFAE/2022/003 funded by MCIN and the European Union NextGenerationEU (PRTR-C17.I1), and by the European Horizon Europe staff exchange (SE) programme HORIZON-MSCA-2021-SE-01 Grant No. NewFunFiCO-101086251. 
ZD is grateful for support from the CIERA Board of Visitors Research Assistant Professorship. 
ACG is supported by the Dutch Research Council (NWO) with project number VI.Veni.212.047. 
DEH is supported by NSF grants AST-2006645 and PHY-2110507, as well as by the Kavli Institute for Cosmological Physics through an endowment from the Kavli Foundation.
DK was supported by the Spanish Agencia Estatal de Investigación grants CNS2022-135440 and PID2022-138626NB-I00, funded by MICIU/AEI/10.13039/501100011033 and the ERDF/EU; and the Comunitat Autònoma de les Illes Balears through the Servei de Recerca i Desenvolupament and the Conselleria d'Educació i Universitats with funds from the Tourist Stay Tax Law (PDR2020/11 - ITS2017-006), from the European Union - NextGenerationEU/PRTR-C17.I1 (SINCO2022/6719) and from the European Union - European Regional Development Fund (ERDF) (SINCO2022/18146).

\bibliographystyle{aasjournal}
\bibliography{sample}

\newpage 

\section*{Author Bios}

\textbf{Hannah Middleton} is a research scientist at the Institute for Gravitational Wave Astronomy at the University of Birmingham, UK. She works on gravitational wave data analysis and is also a member of the LISA ground segment team (a future space-based gravitational wave observatory). She is enthusiastic about public engagement. She has worked on exhibits, art+science collaborations, and is the current Editor-in-Chief of the LIGO Magazine (\url{www.ligo.org/magazine}). \\
\url{https://orcid.org/0000-0001-5532-3622}
\vspace{1em} 

\noindent
\textbf{Christopher P L Berry} is a Senior Lecturer with the Institute of Gravitational Research, University of Glasgow, and a Visiting Scholar with the Center for Interdisciplinary Exploration \& Research in Astrophysics, Northwestern University. He primarily works on gravitational-wave data analysis and understanding the astrophysics of black holes. Christopher has created a variety of public-engagement resources (from graphics to blog posts), helps to coordinate LIGO social media, and is part of the team behind the Gravity Spy community-science project. \\
\url{https://orcid.org/0000-0003-3870-7215}
\vspace{1em} 

\noindent 
\textbf{Nicolas Arnaud} is a CNRS/IN2P3 researcher in France. He is a member of the Virgo Collaboration, working at the interface between instrument science and data analysis. Nicolas has been involved in outreach and education activities for several years, both in France and for the experiments he has worked on. He was the Virgo outreach coordinator at the time of the first detections of gravitational waves and is the French representative to the IPPOG Collaboration.\\
\url{https://orcid.org/0000-0001-6589-8673} 
\vspace{1em} 

\noindent
\textbf{David Blair} is leader of the Einstein-First schools science education project that has created a comprehensive school Einsteinian science curriculum called Eight Steps to Einstein’s Universe, for students from ages 7–15. Starting with toys and songs it coherently builds an intuitive understanding of modern science from atoms and photons, to topics including quantum measurement, cosmology, black holes, gravitational waves and climate science.
\url{https://orcid.org/0000-0002-1501-2405}
\vspace{1em} 

\noindent
\textbf{Jacqueline Bondell} is the Senior Education and Outreach Manager for both the ARC Centre of Excellence (CoE) for Dark Matter Particle Physics and the ARC CoE for Gravitational Wave Discovery (OzGrav). She develops educational content for outreach programs and school incursions, focusing on incorporating innovative technology and science content into curriculum-aligned educational opportunities for students and teachers.\\
\url{https://orcid.org/0000-0002-8089-5662}
\vspace{1em} 

\noindent
\textbf{Alice Bonino} is a PhD student at the University of Birmingham (United Kingdom) working on modeling gravitational waves and numerical simulations of binary black hole mergers. She is also passionate about music and art and coordinated the project Into the Quadrivium with the goal of forging a deep connection between music, art and science while making gravitational waves more accessible to the audience.\\
\url{https://orcid.org/0000-0001-6502-284X}
\vspace{1em} 

\noindent
\textbf{Nicolas Bonne} is a blind astronomer and Public Engagement and Outreach Fellow based at the University of Portsmouth. He leads the Tactile Universe public engagement project which is developing free multi-sensory resources to give blind and vision impaired people more accessible ways to engage with current topics in astronomy.\\
\url{https://orcid.org/0000-0001-9569-8808}
\vspace{1em} 

\noindent
\textbf{Debarati Chatterjee} is Associate Professor at the Inter-University Centre for Astronomy and Astrophysics in Pune, India and the current chair of Education and Public Outreach for the LIGO-India project. Her expertise is in developing theoretical models of neutron stars and searching for signatures of its interior composition in gravitational waves. She is also passionate about communicating science in different languages.\\
\url{https://orcid.org/0000-0002-0995-2329}
\vspace{1em} 

\noindent
\textbf{Sylvain Chaty}, professor, vice-president of culture, science and society at Paris Cité University, is an astrophysicist in the AstroParticle and Cosmology (APC) laboratory. His research focuses on stellar evolution of binary systems, hosting neutron stars and black holes, from formation up to merging, leading to gravitational wave emission. Member of the international collaborations LIGO--Virgo--KAGRA and CTA (high energy astronomical observatory), and passionate about transmission and dissemination of knowledge, he currently is the outreach responsible for the Virgo group of APC.\\
\url{https://orcid.org/0000-0002-5769-8601}
\vspace{1em} 

\noindent
\textbf{Storm Colloms} is a PhD student at the University of Glasgow researching the astrophysical origins of binary black holes detected with gravitational waves. They are also an editor and illustrator for the LIGO Magazine, curator of the Humans of LIGO blog, and author and editor for Astrobites.\\
\url{https://orcid.org/0009-0009-9828-3646}
\vspace{1em} 

\noindent
\textbf{Lynn Cominsky} is an award-winning Professor in the Physics and Astronomy Department at Sonoma State University, and the director of EdEon STEM Learning which develops educational materials for NASA, NSF and the US Department of Education. She also co-chairs the Formal Education Working group for the LIGO Scientific Collaboration, and has led the creation of two online courses for community college instructors, an educator guide for secondary instructors and a video explaining how LIGO operates.\\
\url{https://orcid.org/0000-0003-2073-1065}
\vspace{1em} 

\noindent
\textbf{Livia Conti} is senior researcher at Istituto Nazionale di Fisica Nucleare in Padova, Italy. As an experimental physicists, she has focused mainly at developing gravitational wave detectors, from the old style resonant bars to modern interferometers. Presently her major topic is mitigating noise due to scattered light. She is active in science dissemination, with special care for inclusion and accessibility.\\
\url{https://orcid.org/0000-0003-2731-2656}
\vspace{1em} 

\noindent
\textbf{Isabel Cordero-Carrión} is Associate Professor in the Faculty of Mathematics of the University of Valencia, Spain. She has been a member of the Virgo Collaboration since 2016 and is the current Virgo outreach coordinator. Her research lines focus on applied mathematics and astrophysics, with special interest in the development of numerical methods for partial differential equations, numerical relativity and gravitational waves.\\
\url{https://orcid.org/0000-0002-1985-1361}
\vspace{1em} 

\noindent
\textbf{Robert Coyne} is an associate teaching professor of physics at the University of Rhode Island, and the current chair of the Communications and Education division of the LIGO Scientific Collaboration. His research interests include multi-messenger studies of gamma-ray bursts. He is passionate about education and public outreach, having been involved in both science communication as well as physics education for over a decade.\\
\url{https://orcid.org/0000-0002-5243-5917}
\vspace{1em} 

\noindent
\textbf{Zoheyr Doctor} studies black holes and neutron stars through detections of gravitational waves. He studies how these dead stars are formed and under what circumstances they may collide with each other, releasing a powerful burst of gravitational waves. To explore these topics, Zoheyr makes frequent use of statistical and machine-learning techniques, and enjoys bringing together astrophysics and data science.\\
\url{https://orcid.org/0000-0002-2077-4914}
\vspace{1em}

\noindent
\textbf{Andreas Freise} is a Professor of Gravitational Wave Physics at the Vrije Universiteit Amsterdam and Nikhef. His work focuses on instrument science and interferometer design, from GEO 600 to Virgo, LIGO and the Einstein Telescope. He founded the LIGO Magazine and started Laser Labs Games CIC to make and publish mobile apps for LVK outreach.\\
\url{https://orcid.org/0000-0001-6586-9901}
\vspace{1em} 

\noindent
\textbf{Aaron Geller} studies how stars and planets are born and how they change with time, both through observations and numerical simulations. In addition, he develops visualizations of his and others’ work for both research and outreach purposes, and for use at Northwestern, the Adler Planetarium, within classroom lessons, and for the general astronomy enthusiast.\\
\url{https://orcid.org/0000-0002-3881-9332}
\vspace{1em} 

\noindent
\textbf{Anna C Green} is a postdoctoral fellow at Nikhef, Amsterdam, Netherlands. She studies instrumentation and interferometer design for ground-based gravitational-wave detectors, splitting her time between LIGO, Virgo and Einstein Telescope Collaborations. She is currently Deputy Editor-in-Chief of the LIGO Magazine, and has developed exhibits, apps, and educational materials to engage both new researchers and the general public.\\
\url{https://orcid.org/0000-0002-6287-8746}
\vspace{1em} 

\noindent
\textbf{Jen Gupta} is an Associate Professor in Public Engagement and Outreach at the Institute of Cosmology and Gravitation, University of Portsmouth, and a founding member of the Tactile Universe project.\\
\url{https://orcid.org/0000-0001-9799-4459}
\vspace{1em} 

\noindent
\textbf{Daniel Holz} is a Professor at the University of Chicago, working on gravitational-wave astrophysics and cosmology. He is Chair of the Science and Security Board of the Bulletin of the Atomic Scientists, and in this role helps set the time of the Doomsday Clock. Holz is also founding director of the UChicago Existential Risk Laboratory (XLab), an interdisciplinary effort focused on understanding and mitigating existential risks, including nuclear war, climate change, and AI-fueled disinformation.\\
\url{http://orcid.org/0000-0002-0175-5064}
\vspace{1em} 

\noindent
\textbf{William Katzman} leads the LIGO Science Education Center team as a LIGO Laboratory staff member and Caltech employee. William has been an LSC member since 2009, specializing in education and public outreach.\\
\url{https://orcid.org/0009-0009-6748-754X}
\vspace{1em} 

\noindent
\textbf{Jyoti Kaur} is a physics educator whose research focuses on teaching modern science to school students. She is passionate about STEM education. For over a decade, she has collaborated with schools to design and implement innovative teaching methods that help students gain a deeper understanding and appreciation of modern science. She is currently working with her team to expand the Einstein-First program, aiming to introduce Einsteinian physics to students across Australia and globally.\\
\url{https://orcid.org/0000-0002-3245-5240}
\vspace{1em} 

\noindent
\textbf{David Keitel} works in the LSC since 2011 on gravitational-wave data analysis for a variety of sources. He is a faculty member at the University of the Balearic Islands in Spain. He is currently co-chair of the LVK continuous waves working group, which searches for long-duration gravitational-wave signals from spinning neutron stars, and has managed the LVK science summaries for several years.\\
\url{https://orcid.org/0000-0002-2824-626X}
\vspace{1em} 

\noindent
\textbf{Joey Shapiro Key} is an Associate Professor of Physics at the University of Washington Bothell contributing to data analysis development for gravitational wave observatories. She served as the chair of the Education and Public Outreach committee for the LIGO Scientific Collaboration during the exciting time of the detection and announcement of the first gravitational wave signals. \\
\url{https://orcid.org/0000-0003-0123-7600}
\vspace{1em} 

\noindent
\textbf{Nutsinee Kijbunchoo} is a postdoc at the University of Adelaide. Her current work focuses on investigating the crystal degradations inside of LIGO squeezers. In her spare time Nutsinee likes to play games, drink, and draw things. She is also the author of the Antimatterwebcomics.com. The comic features anything from space cats to students' mental health. Sometimes the comic illustrates collaboration’s hot topics in a very politically incorrect way.\\
\url{https://orchid.org/0000-0002-2874-1228}
\vspace{1em} 

\noindent
\textbf{Carl Knox} joined OzGrav, the Australian Centre of Excellence for Gravitational Wave Discovery, in 2017. He currently holds the title of Creative Technologist and Scientific Visualization Specialist. Carl specialises in data visualisation and artistic representations of astronomical phenomena, including gravitational waves. He uses various mediums, such as 3D animation, graphic design, and real-time generative techniques, to create artworks for media releases, public outreach, and experimental immersive and interactive experiences.
\vspace{1em} 

\noindent
\textbf{Coleman Krawczyk} is a Senior Research Software Engineer and technical lead for the Tactile Universe project at the University of Portsmouth.  He has designed the project’s signature 3D printable models alongside software to make it easier for researchers, educators and communicators to make their own models.\\
\url{https://orcid.org/0000-0001-9233-2341}
\vspace{1em} 

\noindent
\textbf{Ryan N Lang} is a Lecturer in the Concourse Program at the Massachusetts Institute of Technology.  He has worked on LIGO outreach since the first gravitational-wave detection announcement in February 2016, focusing most recently on social media and the \url{questions@ligo.org} service for answering scientific questions from the general public.\\
\url{https://orcid.org/0000-0002-4804-5537}
\vspace{1em} 

\noindent
\textbf{Shane L Larson} is a research professor of physics at Northwestern University, where he is Associate Director of CIERA (Center for Interdisciplinary Exploration and Research in Astrophysics). He works across the gravitational wave spectrum, specializing in compact binaries and the galaxy for both LIGO, and the forth-coming space-based observatory LISA. He teaches a broad range of introductory physics and astronomy, and also teaches for Northwestern’s Prison Education Program. He is an active public science speaker and blogger.\\
\url{https://orcid.org/0000-0001-7559-3902}
\vspace{1em} 

\noindent
\textbf{Susanne Milde} is a communications expert affiliated with the AEI. She is a member of the LIGO Scientific Collaboration, leads the LIGO--Virgo--KAGRA Media Communications Team, the Communications \& Education Group in the Einstein Telescope Consortium and coordinates the LISA Consortium media activities. She is specialized in international science communications and passionate about communicating fundamental science to the public. An ecologist by education she has worked in university, politics and industry before founding Milde International Science Communications.
\vspace{1em} 

\noindent
\textbf{Vincenzo Napolano} is the head of communication at the European Gravitational Observatory (Pisa), where he is in charge of national and international communication related to the Virgo gravitational wave antenna. He has been a scientific communicator at the Italian Institute of Nuclear Physics (INFN). He has curated many projects and installations at the intersection between new technologies, artistic languages and communication of science, events and exhibitions in collaboration with Foundations, Art and Science Museums and Festivals in Europe and abroad.
\vspace{1em} 

\noindent
\textbf{Chris North} is a Reader and Head of Public Engagement at the School of Physics and Astronomy at Cardiff University. His activities include data visualisation and educational workshops across a range of areas, but particularly astronomy and gravitational waves. He has developed a number of schools workshops and interactive website.
\vspace{1em} 

\noindent
\textbf{Sascha Rieger} is a science communicator affiliated with the Max Planck Institute for Gravitational Physics Albert Einstein Institute, in the gravitational wave research field for almost 25 years. He has worked on any number of movies, animations, graphics, exhibitions, and even more publications - reports, grant applications, brochures, flyers, posters among others. These days he is busy with the LIGO Magazine, one of the public outreach leads in the LISA Consortium, and doing Einstein Telescope communication whenever he finds the time.
\vspace{1em} 

\noindent
\textbf{Giada Rossi} is a science communicator based in Italy, specializing in institutional communication in the field of physics and astronomy. After graduating in Astronomy from the University of Bologna, she attended SISSA's Master in Science Communication. She works in the communication unit at the European Gravitational Observatory (EGO), the institution that hosts the Virgo gravitational wave detector in Pisa, Italy.
\vspace{1em} 

\noindent
\textbf{Hisaaki Shinkai} is a professor at the Faculty of Information Science and Technology, Osaka Institute of Technology, Japan. He works on theoretical general relativity and cultural studies of astronomy. He led the KAGRA Scientific Congress as a board chair 2017–2021, and started the KAGRA Education and Public Outreach group. He organizes circular exhibitions on black-hole and gravitational waves at Japan Science Museums 2025–2028, and has published 10+ books for students and for general readers.\\
\url{https://orcid.org/0000-0003-1082-2844}
\vspace{1em} 

\noindent
\textbf{Aurore Simonnet} is the scientific illustrator for the EdEon STEM Learning team at Sonoma State University. She adds a creative and imaginative flare to physical science and technical illustrations used for various education and public outreach programs (LIGO among others!). From diagrams to astronomical illustrations, her skills in graphic design and technical drawing contribute a diverse blending of scientific and artistic perspective.
\vspace{1em} 

\noindent
\textbf{Andrew Spencer} is a Lecturer at the Institute for Gravitational Research at the University of Glasgow, his research focus is laser interferometry and technology development for future gravitational wave detectors. He has a special interest in outreach resource development, making education and communication resources for classrooms, science centres and festival events in addition to working on crossover projects that span science and creative arts.\\
\url{https://orcid.org/0000-0003-4418-3366}

\end{document}